\documentclass[11pt]{article}

\usepackage[letterpaper,top=2cm,bottom=2cm,left=3cm,right=3cm,marginparwidth=1.75cm]{geometry}

\usepackage{amsmath}
\usepackage{graphicx}
\usepackage{authblk}
\usepackage[colorlinks=true, allcolors=blue]{hyperref}
\usepackage{cleveref}
\usepackage{lipsum}% Just for this example
\usepackage{graphicx,caption}
\usepackage{caption}

\newcommand{\iu}{\mathrm{i}\mkern1mu}
\renewcommand\vec{\mathbf}
\usepackage{braket}

\title{Retrieval of phase information from low-dose electron microscopy experiments: are we at the limit yet?} 

\author[1,2]{\small Francisco Vega Ibáñez}
\author[1,2,*]{\small Johan Verbeeck}
\affil[1]{\footnotesize University of Antwerp, EMAT, Groenenborgerlaan 171, 2020, Antwerp, Belgium}
\affil[2]{\footnotesize NANOlab Center of Excellence, University of Antwerp, Groenenborgerlaan 171, 2020 Antwerp, Belgium}
\affil[*]{\footnotesize Corresponding Author: Jo Verbeeck \href{mailto:jo.verbeeck@uantwerp.be}{jo.verbeeck@uantwerp.be}}
\date{}                     %% if you don't need date to appear
\providecommand{\keywords}[1]{\textbf{\textit{Keywords: }} #1}
\usepackage{cleveref}

\begin{document}
\maketitle

\abstract{The challenge of imaging low-density objects in an electron microscope without causing beam damage is significant in modern TEM. This is especially true for life science imaging, where the sample, rather than the instrument, still determines the resolution limit. Here, we explore whether we have to accept this or can progress further in this area. To do this, we use numerical simulations to see how much information we can obtain from a weak phase object at different electron doses. Starting from a model with four phase values, we compare Zernike phase contrast with measuring diffracted intensity under multiple random phase illuminations to solve the inverse problem. Our simulations have shown that diffraction-based methods perform better than the Zernike method, as we have found and addressed a normalization issue that, in some other studies, led to an overly optimistic representation of the Zernike setup. We further validated this using more realistic 2D objects and found that random phase illuminated diffraction can be up to five times more efficient than an ideal Zernike implementation. These findings suggest that diffraction-based methods could be a promising approach for imaging beam-sensitive materials and that current low-dose imaging methods are not yet at the quantum limit.}
\\
\keywords{Zernike phase contrast, phase retrieval, inverse problem, ptychography, multiplex holography, phase plates}

\maketitle

\section{Introduction}
In this work, we aim to revisit the longstanding issue of phase reconstruction in transmission electron microscopy (TEM) \cite{drenth1975problem,coene1992phase,Fienup1982PhaseComparison,mccallum1992two,rodenburg1993experimental,mccallum1993simultaneous} and examine it from the perspective of information transfer. 
In electron microscopy, we detect electron events that are quantum-mechanically linked to their wavefronts' probability density (square modulus). The problem arises from the loss of the phase during the detection process, which significantly restricts the information obtainable from an electron microscope experiment. This issue is particularly challenging in electron diffraction experiments, as it hinders the extraction of the projected periodic potential of a crystal. Moreover, it is also highly relevant in imaging non-periodic thin objects in TEM, where the object's projected density information is predominantly encoded in the phase profile imparted on the coherent plane wave illumination. Recent attempts to apply diffraction-based imaging, e.g., to viruses \cite{Zhou2020Low-dosePtychography} or in single particle analysis \cite{Pei2023CryogenicTransfer} show great potential and
are accompanied by promising simulation studies \cite{Pelz2017Low-doseOptimization,Leidl2023DynamicalMicroscopy,Mao2024Multi-Convergence-AngleResolution}.

We will use the toolset of parameter estimation, which has shed light on similar problems in TEM, like investigating point resolution in the presence of noise \cite{Bettens1999Model-basedStatistics}, the advantage of a monochromator on the spatial resolution in TEM \cite{denDekker2001DoesHRTEM}, determining the precision of measuring atomic positions from exit waves \cite{DeBacker2011HighReconstruction}, or even determining elemental concentrations from electron energy loss experiments \cite{Verbeeck2004ModelSpectra,Https://github.com/joverbee/pyEELSMODEL}.
The issue of phase retrieval under dose-limited conditions has sparked significant debate within the scientific community. Egerton et al. conducted groundbreaking research to assess the instrument's limitations \cite{egerton2007limits} and evaluated different commonly used TEM and STEM imaging methods on beam-sensitive specimens \cite{egerton2013control}. 
In the following years, more theory was incorporated to assess the efficiency of phase retrieval by incorporating robust mathematical concepts such as the Fischer Information (FI) and Cramér-Rao Lower Bound (CRLB) \cite{bouchet2021fundamental}. Based on these mathematical concepts, Koppell and Kasevich constructed a function to assess the inherent frequency transfer of the imaging system \cite{koppell2021information}. More recently, Dwyer and Paganin directly compared Zernike Phase Contrast (ZPC) and 4D-STEM with a phase-structured illumination. 
All this notable work has paved the way and opened the debate for a more comprehensive assessment of phase retrieval in the TEM, with the general conclusion that ZPC seemed the best method to maximize information transfer. Ultimately, this conclusion has put limits on the hope for ptychographic methods to create a breakthrough in low-dose phase imaging \cite{Fienup1982PhaseComparison,Maiden2009AnImaging,Faulkner2004MovableAlgorithm,Nellist1995ResolutionMicroscopy,Yang2015EfficientConditions,Parvizi2015AEquation}. Here, we revisit this problem by conducting a series of numerical exercises. Furthermore, we will carefully consider the normalization conditions to enable a fair comparison between image-based reconstruction using ZPC and phase retrieval through diffraction-based recording. We demonstrate that at least under the idealised conditions considered here, significant improvement over ZPC in low dose phase imaging is possible with diffraction based detection.

\section{Setup}

\begin{figure}[b!]
    \centering
\includegraphics[width=.88\linewidth]{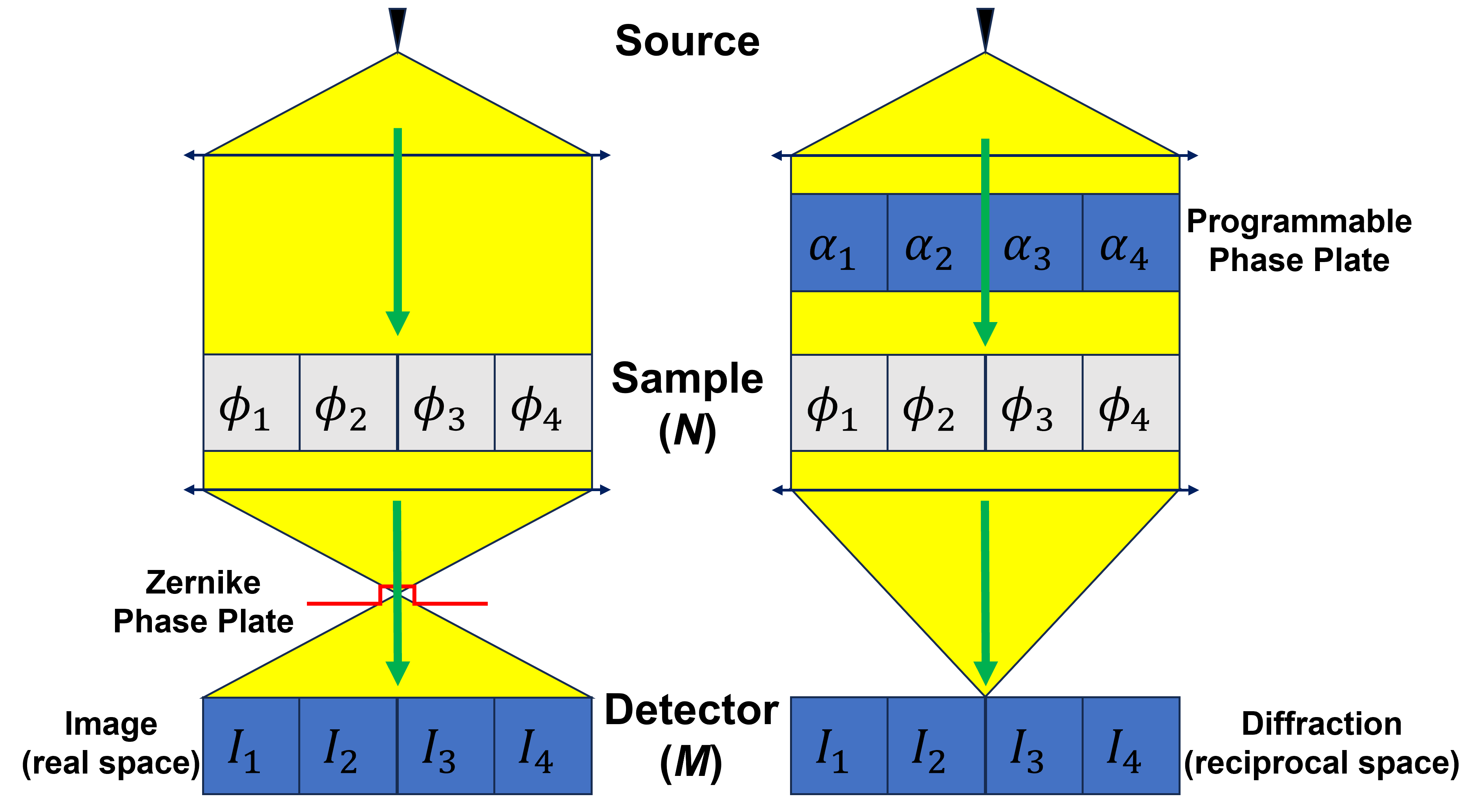}
    \caption{Sketch of the setup. The aim is to measure the $N$ phases of the unknown object by illuminating it with a coherent electron wave and detecting the arrival of electrons with an ideal detector consisting of $M$ independent pixels. The left-side setup shows the configuration with an ideal ZPP, while the right-side setup describes the detection in the diffraction plane with a programmable phase input wave. We will compare the performance of both setups in terms of phase error on the estimate as a function of the amount of electrons we have available in the experiment.}
    \label{fig:sketch}
\end{figure}

In the following section, we want to present this process as a type of \textit{game} where the sample is imagined to contain a hidden message made up of $N$ phase values, denoted as $\phi_i$ in Fig. \ref{fig:sketch}. We illuminate the sample with an electron wave and observe the outcome of this interaction on $M$ ideal electron detectors, labeled as $I_i$ in Fig. \ref{fig:sketch}. From this simple setup, two natural questions arise:

\begin{itemize}
    \item How many electrons do we need to \textit{fire} onto the sample to obtain the secret message at the required precision and accuracy?
    \item How can the experiment be set up to achieve the best precision and accuracy with the fewest electrons and, thus, the least beam damage? 
\end{itemize}

These questions are fundamental in modern electron microscopy, as the resolution of EM is in many practical cases limited by beam damage and not anymore by the instrument \cite{glaeser1971limitations,chen2008dose,muller2019comparison,nakane2020single,Pei2023CryogenicTransfer,chari2023prospects,kuccukouglu2024low,Leidl2024InfluenceFlow}. This means we must either learn new techniques to limit beam damage or utilize the most efficient imaging methods to maximize the use of the electron dose the sample can withstand (preferably a combination of both).

In this paper, we will avoid all complications regarding the scattering that happens with the sample, details of imperfect optical systems \cite{gonsalves1982phase,fienup1993phase}, propagation effects \cite{liu2009influence,robert2022dynamical}, multiple scattering in the sample \cite{maiden2012ptychographic,chen2020multi,ren2020multiple,chen2024imaging}, inelastic scattering \cite{yoshioka1957effect,muller1995delocalization,dwyer2005multislice,dwyer2008multiple,allen2015modelling,brown2018structure,beyer2020influence,robert2022dynamical,diederichs2024exact}, partial coherence \cite{nellist1994beyond,gureyev2006linear,martin2006spatial,thibault2013reconstructing,oxley2020importance,diederichs2024exact}, and details of the algorithmic implementation \cite{Fannjiang2012PhaseIllumination,elser2018benchmark} to gain some clarity on how far we are from fundamental limits.

We start with a conceptual exercise to estimate four hidden phases as sketched in Fig. \ref{fig:sketch}. We compare two typical setups:
On the one hand, we use a Zernike Phase Plate (ZPP) for phase contrast imaging, which is commonly considered the golden standard in real space phase imaging and is used extensively in, e.g., life science imaging \cite{Zernike1942PhaseObjects,Zernike1942PhaseII,Danev2001TransmissionPlate,danev2008single}. The benefit of this method is that it results directly in an image of the sample with a contrast that relates approximately linearly to the phase shift, which is proportional to the projected electrostatic potential of the thin sample in a TEM.
 
On the other hand, we can detect the scattered electrons in the diffraction plane as is commonly done to investigate symmetries and periodicity in crystals. This pattern also encodes the information of the specimen albeit in a different way and requiring some inverse algorithm to link the recorded intensities to the projected sample potential we are interested in.

In either case, retrieving the \textit{absolute phase} will be impossible as we have no unperturbed reference beam to compare. Due to this lack of a reference beam, only three of the four unknown phases are independent, somewhat simplifying the problem from $N$ to $(N-1)$ unknowns represented as $\phi_N=-\sum_1^{N-1} \phi_i$.

Because either a translation or an inversion of the object leaves the diffraction intensities unchanged, we have a good chance of ending up with a wrong guess of the secret sample for the diffraction-based setup \cite{guizar2012understanding,tolimieri2012mathematics}. A typical way to solve this is to oversample the diffraction plane ($M > N$), which stabilizes the solution at the expense of requiring more detector pixels.
Another way to proceed is by introducing an amplitude \cite{Allars2021EfficientDiffuser,abregana2022phase,you2023lorentz} or phase \cite{Yu2023QuantumElectrons,VegaIbanez2023CanTEM,Verbeeck2018DemonstrationElectrons} modulator capable of encoding the electron wavefront for $N_{config.}$ sets of conditions. 
Here, we focus on pure phase modulation without delving into the details of how to create such a programmable phase modulator \cite{VegaIbanez2023CanTEM,Yu2023QuantumElectrons} and simply assume it to be perfect, as we did for ZPC.

Suppose we choose a number of Random Phase Illumination (RPI) conditions $N_{RPI}$, and we solve the inverse problem by taking into account the $N_{RPI}$ independent measurements to resolve one unique estimate of the object phase. We now obtain some robustness against inversion and translation since the extra configurations yield $M\times N_{RPI}$ measurement points (far more than the $N$ unknown phases we want to recover).

In order to implement this scheme, we use a nonlinear Maximum Likelihood (ML) fitting algorithm with $(N-1)$ unknown phases and a likelihood function assuming Poisson counting noise that describes how likely it is that a given experimental realization of $M\times N_{RPI}$ diffraction intensities could have been produced when assuming a given set of $(N-1)$ sample phases \cite{Barlow1991ASciences,Cramer1946MathematicalStatistics}.
This iterative nonlinear fitting process is significantly slower than the more common Gerchberg-Saxton (GS) \cite{Gerchberg1972APictures} algorithm but allows the correct treatment of the Poisson statistics and obtaining estimates for the connection between phase errors and the counting statistics through the use of the Cram\'er-Rao lower bound \cite{Rao1945InformationParameters, Cramer1946MathematicalStatistics}. We can then use this to compare the behavior of the GS algorithm to ML prediction to convince ourselves that it approaches the same fundamental limit while providing a significant speed-up needed for realistic image sizes.

\section{Recovering the phase and estimating its precision}
Our objective is to accurately determine the unknown phase from either a real-space or diffraction-space intensity recording. We will use the ZPC method for real space as a standard, and in the case of the diffraction experiment, we will need to solve the inversion problem. To do so, we will utilize parameter estimation to understand its statistical properties and, later, use a GS algorithm that can approach these while providing a significant numerical speed advantage.

\subsection{Zernike Phase Contrast}
In Zernike Phase Contrast \cite{Zernike1942PhaseObjects,Zernike1942PhaseII}, a phase plate is placed in the back focal plane of the objective aperture, which shifts only the low-frequency component of the wave by $\pi/2$. As a result, the image contrast now reveals the phase of the object. In appendix \ref{ap:appzernike}, we derive the relationship between phase and image contrast, paying attention to proper normalization (see equation \ref{eq:zernikephasecorrected}). This aspect has remained under-illuminated but will be critically important when studying the effect of counting noise:
\begin{equation}
\phi(\vec{r})=\frac{I_z(\vec{r})-1+2C^2-2C^4}{2C^2}
\label{eq:z_phase}
\end{equation}
With $I_Z$, the recorded intensity scaled such that if the ZPP were removed, there would be an average of 1 electron per pixel.
The factor $C=Rk_Z$ is given as the radius of the illumination disc $R$ on the sample times the size of the central phase shifting area $k_Z$ in reciprocal units, and the formula is valid only for $C \leq 1/2$ as we will lose low-frequency information for higher values of $C$.
In the derivation, we assumed a Heavyside pi function for the phase profile, which gives the most faithful phase recovery (but may be harder to achieve in practice).
The critical thing to note here is that this formula only agrees with the conventional ZPC for $C=1$. However, in this case, the formula is not valid, as important low-frequency information would be missing across the illuminated area because the reference wave generated by the central Zernike phase discontinuity would not be homogeneous.
Only if we admit to being interested in a subregion of the illuminated area can we recover the conventional formula at $C=1$. It is important to stress here that, in such a case, we create a situation that can no longer act as a fair comparison with the diffraction setup for the following reasons:
\begin{itemize}
\item{We are illuminating and damaging areas of the sample outside the field of view. We might need those areas later on.}
\item{We use electrons outside the field of view which will help to create a reference beam inside the field of view. This effectively creates a setup similar to an off-axis holography experiment \cite{lichte2007electron} and results in an unfair counting of the incoming amount of electrons that is needed per area. If such an external reference wave can be added at no penalty, then this option should also be offered to the diffraction setup, e.g. by assuming that part of the illuminated field of view is known to be constant or of no interest. This would lower the amount of unknowns and increase the precision as well.}
\item{Even if the area of the sample around the region of interest can be considered uninteresting or sacrificial, the electrons hitting there can still cause damage inside the area of interest via delocalized inelastic scattering \cite{egerton_mechanisms_2012,Egerton2017, Jannis2022ReducingProcess, Velazco2022ReducingFindings}}
\end{itemize}
In order to recover all spatial frequencies within the illuminated area without creating an (unfair) implicit reference wave, we choose to take the optimal value of $C=1/2$ for the remainder of this work. With that normalization in mind, we can estimate the standard deviation of the phase error for $N_e$ electrons as:
\begin{eqnarray}
\sigma_{\phi,Z,2D}=\frac{\sqrt{1-2C^2+2C^4}}{2C^2\sqrt{N_e/(N-1)}}\stackrel{C=1/2}{\rightarrow{}} \frac{\sqrt{10}}{2\sqrt{N_e/(N-1)}}
\label{eq:zernikeerrorcorrected}
\end{eqnarray}
Which is $\sqrt{10}\approx 3.16 $ times \emph{higher} than what we would get assuming the wrongly normalized conventional solution.
Note that, for a 1D case, the normalization penalty is less severe:
\begin{eqnarray}
\sigma_{\phi,Z,1D}=\frac{\sqrt{1-2C+2C^2}}{2C^2\sqrt{N_e/(N-1)}}\stackrel{C=1/2}{\rightarrow{}} \frac{\sqrt{2}}{2\sqrt{N_e/(N-1)}}
\label{eq:zernikeerrorcorrected1D}
\end{eqnarray}

To make matters more confusing, it turns out that when numerically implementing ZPC, this aliasing error is often made by wrongly assuming that illumination and sample are periodically repeated until infinity, resulting in a situation that exactly replicates the $C=1$ case without even having the penalty of losing the low-frequency information. This subtle error that we discuss in more detail in appendix \ref{ap:appzernikenumerical} is likely why, in other numerical attempts, the counting noise effect is portrayed too optimistically \cite{koppell2021information,Dwyer2023QuantumMicroscopy}. To help overcome this situation, we include a numerical reference implementation with this paper \cite{verbeeck_2024_13302024}.

\subsection{Inversion via parameter estimation}
Estimating the parameters of a forward model is a simple method for determining the unknown phases from a recorded diffraction pattern \cite{aster2018parameter}.
We can model the object as a discrete pure-phase object:
\begin{eqnarray}
   S_i = e^{\iu \phi_i} 
\end{eqnarray}
Then, we can describe the illumination over this object with a wave function:
\begin{eqnarray}
   \Psi_{i} = A_i e^{\iu \alpha_i} 
\end{eqnarray}
To allow for both amplitude and phase modulation, we can write the exit wave of the object as:
\begin{eqnarray}
   \Psi_{i} = A_i e^{\iu (\phi_i+\alpha_i)} 
\end{eqnarray}
Moreover, we record the diffraction intensities as follows:
\begin{eqnarray}
   I_{model,j} = |\mathcal{F}_j \Psi_{i}|^2 
\end{eqnarray}
With $\mathcal{F}_j$ representing the Fourier transform operator, which projects the wave onto the back focal plane where the detection process occurs.
Now, we can define a maximum likelihood estimator (MLE) based on the assumption that the detection in the diffraction plane is governed by counting noise, and each detector pixel is assumed to be independent. We can write the log-likelihood $l$ \cite{van1989estimation} as: 
\begin{equation}
l\approx  -\sum_j \left(I_{exp,j} \ln(I_{model,j})- I_{model,j}\right)
\end{equation}
With $I_{exp,j}$ and $I_{model,j}$, the experimental and model intensities, respectively.
We can now use a nonlinear fitting algorithm to find the estimates $\tilde{\phi_i}$ of $\phi_i$. 
The ML algorithm has the benefit of delivering estimates with the highest possible precision and without bias for a high number of observations \cite{den2005maximum}, assuming the model and the noise model are correct.
In appendix \ref{ap:appmodelbased}, we derive the Jacobian for this model, which significantly speeds up the iterative optimization routine. Furthermore, we also do the derivation for the Fisher information matrix, which can be inverted to obtain the Cram\'er-Rao Lower Bound (CRLB) \cite{den2005maximum}.

\begin{equation}
\sigma_{\phi,ML} \geq CRLB=\sqrt{\frac{Tr(F^{-1})}{N-1}}
\end{equation}
With Tr(), the Trace operator, and F, the Fisher information matrix.
We can further estimate the CRLB from a more fundamental perspective. We detect in the reciprocal plane where, in principle, information related to the phase of the object is encoded in both amplitude and phase in that plane. As a result, we can, at best, obtain an average of half of the information, as we have assumed no prior information about the object. We could, therefore, assume that the CRLB will be close to the following:
\begin{eqnarray}
    CRLB \approx \frac{\sqrt{2}}{2\sqrt{N_e/(N-1)}}
    \label{eq:crlbestimate}
\end{eqnarray}
This was given as well in Eq. \ref{eq:zernikeerrorcorrected1D} as the correction for the 1D case in a ZPC system with $C=1/2$. However, we will numerically test this idea further on.

The CRLB can predict the best possible precision that can be obtained under the given electron dose. This can then be compared to the actual outcome of a numerical experiment to evaluate whether we can attain the CRLB in practice. It can also serve to compare alternative algorithms like GS to evaluate how close this can approach this limit and check if bias is introduced.

\subsection{Inversion via the Gerchberg-Saxton algorithm}

The Gerchberg-Saxton (GS) algorithm and its variations are iterative methods used to recover phase information from intensity measurements of a complex-valued wave function \cite{Gerchberg1972APictures,yang1994gerchberg,zalevsky1996gerchberg,huang2020generalizing}. 
In our case, starting from a set of intensity recordings in diffraction space $\hat{I} = \{I_1, I_2, \cdots, I_{N_{RPI}}\}$, and their corresponding known illumination patterns in real space $\hat{\Psi} = \{\Psi_1,\Psi_2,\cdots,\Psi_{N_{RPI}}\}$, we aim to retrieve the $(N-1)$ missing phases of an object $\hat{S}$. 
To do this, we can start by guessing our solution $\hat{S}^\star_0$ as a set of $N$ complex numbers with amplitude 1 and random phases $\phi_i \in [-\pi,\pi)$, illuminated by the complex-valued waves from each $N_{RPI}$ measurement in diffraction space. And, in general, for any given iteration:
\begin{equation}
\hat{\varphi}_{out,i}^{\star} = \mathcal{F}[\hat{S}^\star_i\hat{\Psi}] 
\end{equation}
From this, we can impose the constraint of our recording in diffraction space and re-construct our guessed $\hat{\varphi}^{\star}_{out,i+1}$ as:
\begin{equation}
\hat{\varphi}^{\star}_{out,i+1} = \frac{\hat{\varphi}_{out,i}^{\star}}{\|\hat{\varphi}_{out,i}^{\star}\|}\sqrt{\hat{I}} \times \text{Phase}\left(\hat{\varphi}^{\star}_{out,i}\right)
\end{equation}
Where the $\text{Phase}$ operator returns the phase of the given array, respectively. 
From this, we can back-propagate $\hat{\varphi}_{out,i+1}^{\star}$ to real space and obtain a new estimate for the object:
\begin{equation}
\hat{S}_{i+1}^\star = \text{Average}\left(\mathcal{F}^{-1}[\hat{\varphi}_{out,i+1}^\star] \hat{\Psi}_{in}^{\dagger}\right)
\label{eq:update_s}
\end{equation}
With $\hat{\Psi}_{in}^{\dagger}$ being the complex conjugate of the transposed illumination pattern wave set. This process is then carried out iteratively until a set number of iterations or a convergence condition is met.

It's important to note that in Eq. \ref{eq:update_s}, the update takes the average update from each of the  $N_{RPI}$ illuminating patterns. This helps reduce the typical twin and translational artifacts in the Fourier transform when $N_{RPI}>1$. For $N_{RPI}=1$, we will encounter a similar problem to that of ZPC, where oversampling is needed to solve all $N$ phase values in our object.

\section{Numerical exercise for 4 unknown phases (1D)}

\begin{figure}[!ht]
    \centering
    \includegraphics[width=.80\linewidth]{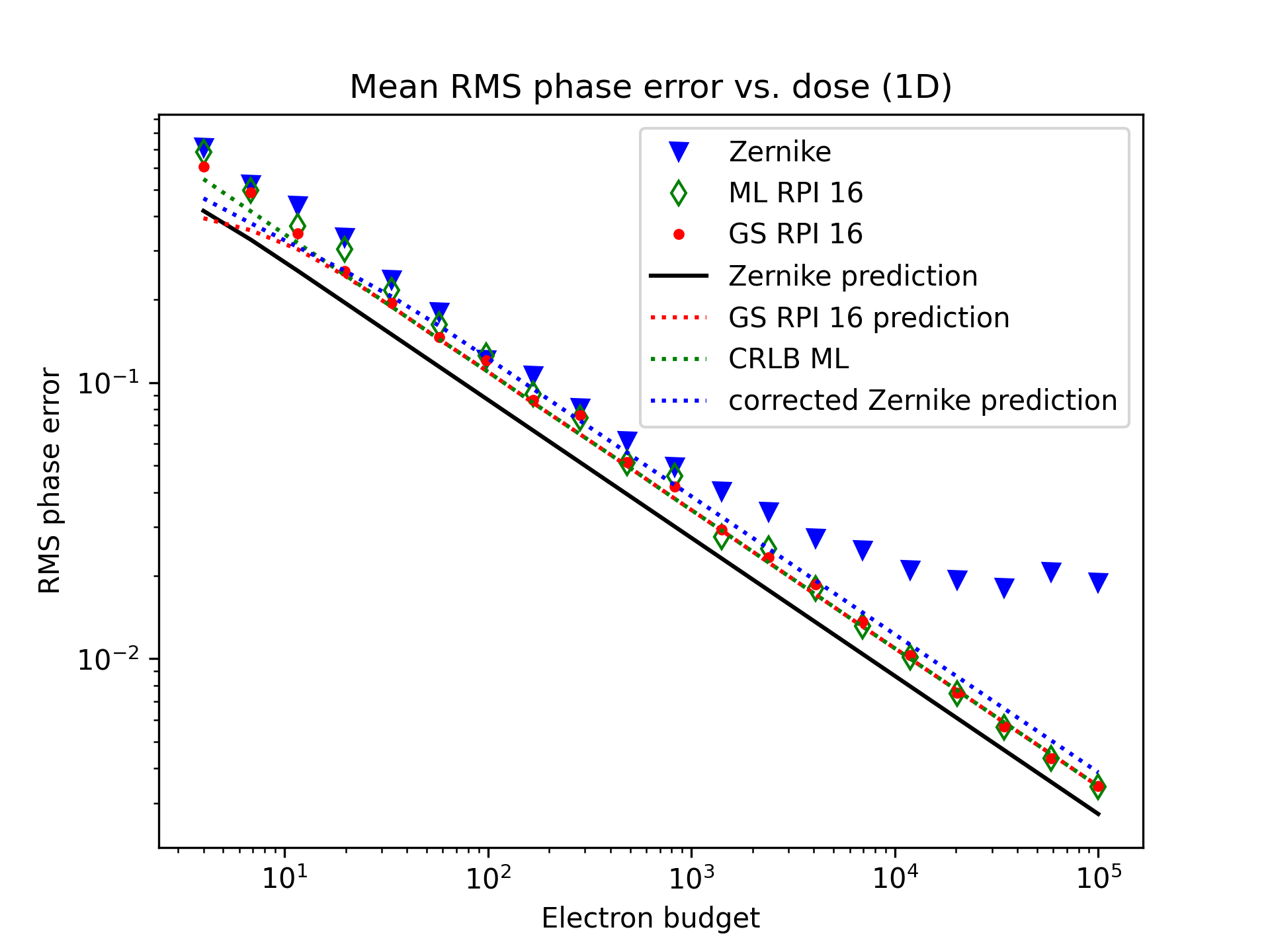}
    \caption{Numerical simulation of the average RMS phase error as a function of total electron dose. The object is considered a random phase object with four unknown phase values and a limited phase range of $\pm\pi/10$ to stay within the linear approximation of the ZPC formula. The simulation is repeated $100$ times for each electron budget with a different random object. Random phase illumination is done with 16 random illumination patterns that stay the same throughout the simulation set. Note how ZPC and RPI's results closely follow the predicted statistical error and show a similar noise performance up to about $N_e=10^4$, where systematic errors due to the linearization of the ZPC formula start to show. Note also how ML RPI and GS RPI perform remarkably similarly, giving confidence in the GS approach for larger systems.}
    \label{fig:error_vs_dose_n4}
\end{figure}

An example of an outcome of this numerical experiment is given in 
Fig. \ref{fig:error_vs_dose_n4} for $N_{RPI}=16$. We note a logarithmic behavior between phase errors as a function of the total number of electrons (as expected). Furthermore, for this 1D case, both RPI and ZPC perform similarly. This performance shows that the corrected error prediction for ZPC in Eq. \ref{eq:zernikeerrorcorrected1D} describes the propagation of the counting noise to the estimated phase quite well.
For very high doses (above $N_e = 10^4$ here), the ZPC phase error does not decrease anymore; this is due to hitting the limit of the linearization of the ZPC formula (Eq. \ref{eq:zernikelinear}), which also depends on the range of phase modulation $\delta$ by the sample (here $\pm \pi/10$). We are, however, more interested in the low-dose performance.
Both the ML and GS results are well predicted by the CRLB, showing that, on average, we approach the statistical limit for 100 repeated experiments per electron dose value. It also shows that our rough estimate in Eq. \ref{eq:crlbestimate} is reasonably accurate (in this case, it coincides with the ZPC corrected prediction) albeit slightly higher than the actual CRLB that is based on the detailed model. The fact that GS attains the CRLB also gives confidence in this much faster algorithm, which will be needed for larger systems later.
We investigated the role of $N_{RPI}$ in the number of random phase illuminations (not shown) and observed that from $N_{RPI} \geq 2$, we converge to the optimal phase error. This convergence confirms the theoretical prediction in \cite{Fannjiang2012PhaseIllumination}. Increasing $N_{RPI}$ leads only to faster convergence, which might benefit larger systems. However, from here on, we will assume $N_{RPI}=16$ unless otherwise noted.
Furthermore, we found that convergence is best for random phases from a uniform distribution between $0$ and $\pi$. However, other phase modulation ranges for the illumination were tested with similar results, provided that the object's phase range is fixed. Other phase patterns were attempted, such as illuminating with only 1 pixel at a random phase between $-\pi/2$ and $+\pi/2$ for each RPI configuration or using an orthogonal set of waves (e.g. a Hadamard basis set \cite{hadamard1893resolution}), both of which also yielded good results.

\section{Numerical exercise for a small 2D case}
\begin{figure}[ht!]
    \centering
    \includegraphics[width=.88\linewidth]{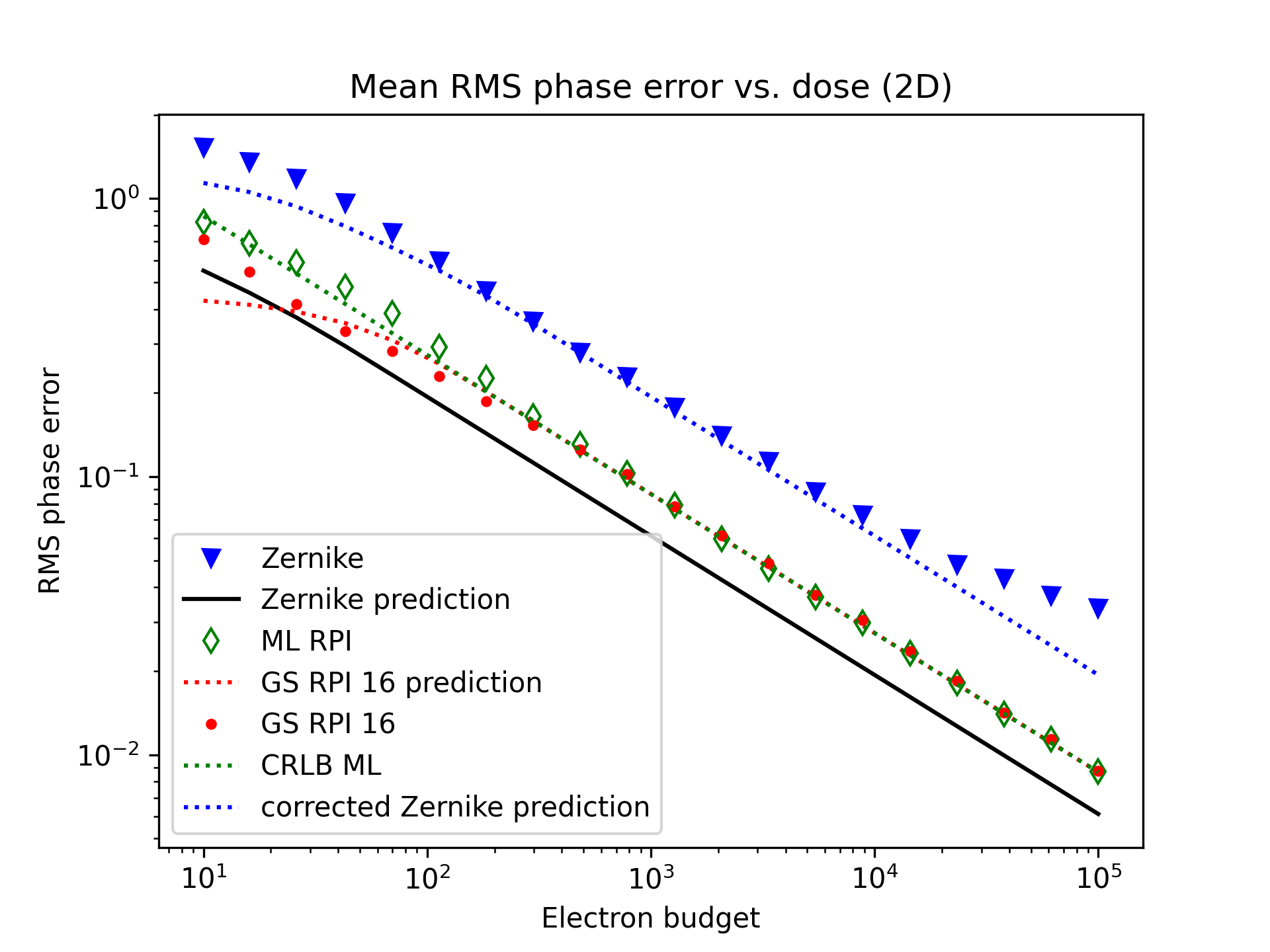}
    \caption{Numerical simulation of a $4\times4$ phase object encoding 100 different realizations of random phase noise with a range of $\pm \pi/10$ for each electron budget. The electron budget is divided over 16 random phase illumination patterns. Note the significant difference with ZPC due to the normalization correction and the close relation between the ML RPI and the much faster GS RPI.}
    \label{fig:4x4}
\end{figure}

Changing to a more common 2D configuration, we attempt a $4\times4$ pixel phase object shown in Fig. \ref{fig:4x4}.
The diffraction-based recording performs better than the corrected ZPC intensity, increasing dose efficiency by $ 5 $. Comparing ML RPI with GS RPI shows that GS attains the CRLB rather well, which gives faith in the algorithm as a faster alternative to ML, allowing for reconstructing larger objects in a reasonable time.
The higher number of unknowns also stabilizes the statistical errors, and the observed behavior for 100 random weak phase object realizations results in an average behavior that closely follows the CRLB and ZPC error predictions. In this case, we barely start appreciating the systematic linearization error for the ZPC setup as the overall error has increased due to the higher number of unknowns, which shifts this point to higher doses.

\section{A more realistic 2D object}
\begin{figure}[ht!]
    \centering
    \includegraphics[width=.88\linewidth]{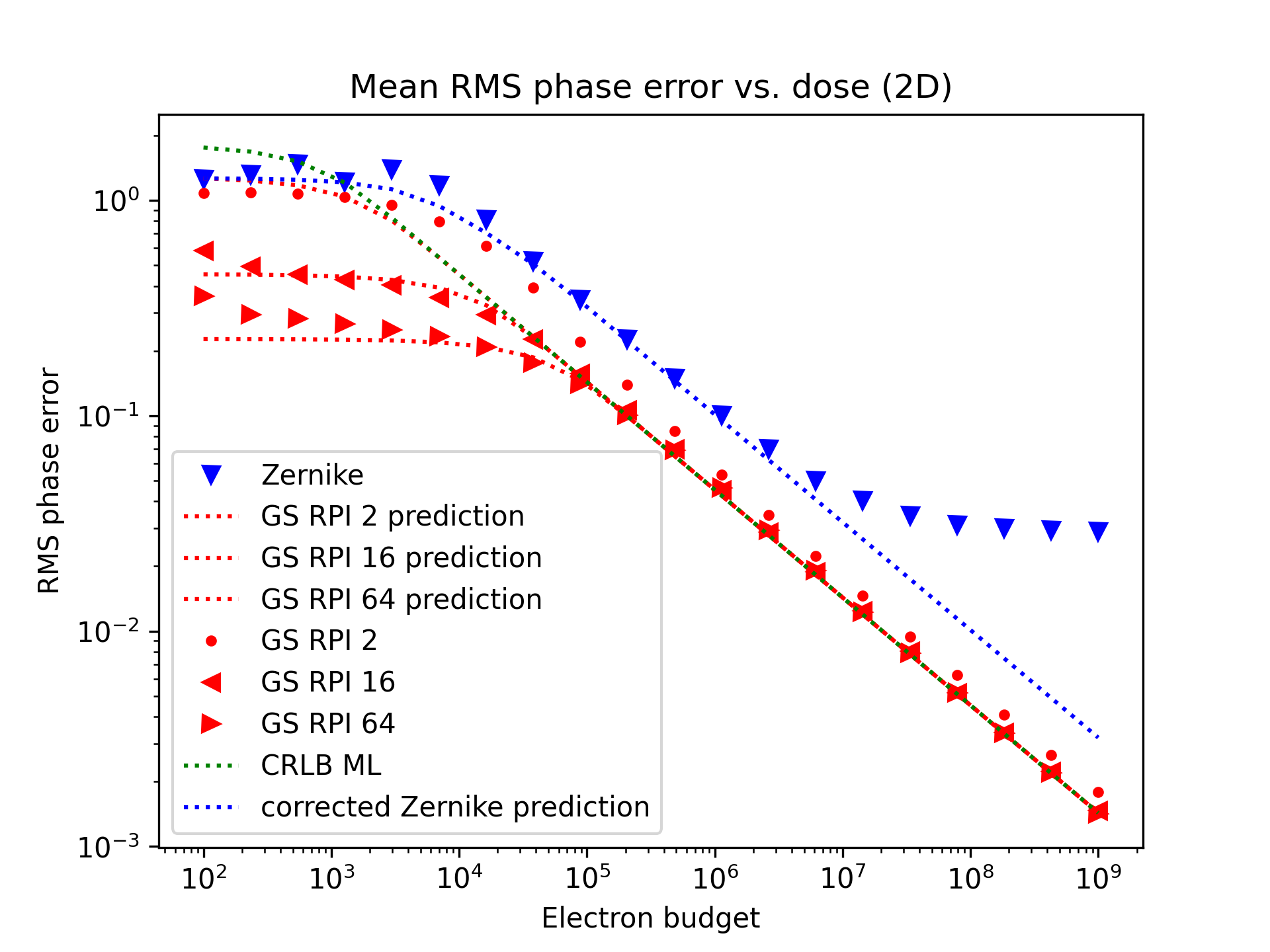}
    \caption{RMS phase error for a single $64\times64$ random phase object with $\pi/10$ phase range, illuminated with $N_{RPI}={2,16,64}$. We note that ZPC and GS closely follow their predicted error behavior as a function of the total dose. We note a remarkable trend at low doses where the phase error shows a plateau depending on $N_{RPI}$, which is a consequence of phase wrapping in the individual $N_{RPI}$ realizations.}
    \label{fig:error64x64}
\end{figure}

\begin{figure}[ht!]
    \centering
    \includegraphics[width=.88\linewidth]{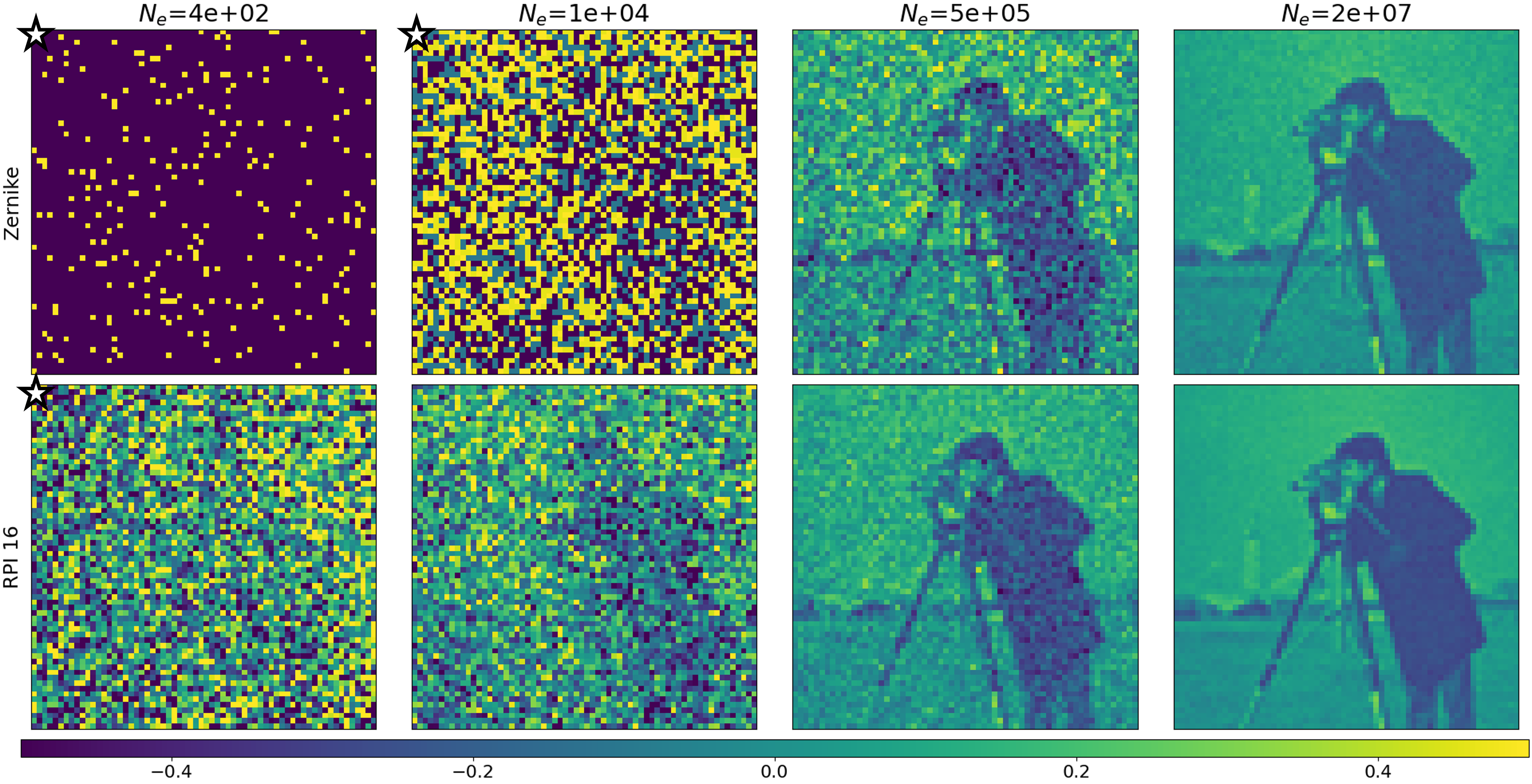}
    \caption{Comparison of ZPC and GS RPI phase retrieval as a function of dose on a more realistic 2D object. At high doses, both methods retrieve the object effectively. The top figures marked with a $\star$ do not adhere to the color bar scaling to aid visualization.}
    \label{fig:cameraman}
\end{figure}

Due to computational limitations, we have decided to use the GS algorithm for larger objects. As observed in previous examples the GS algorithm closely approximates the ML CRLB, in agreement with, e.g., \cite{melnyk2023convergence}.
In Fig. \ref{fig:error64x64}, we display the observed Root Mean Squared (RMS) phase error for a $64\times 64$ random phase object with a $\pm \pi/10$ phase variation. Both methods effectively retrieve the object at high electron doses, but the standard deviation of the phase error remains about $\sqrt{5}$ higher for the correctly sampled ZPC case. We observe a similar systematic error when $N_e \geq 10^7$, which is expected at higher doses compared to the $N=4$ 1D case. This is because we have to estimate approximately 1000 times more phases, so the error requires a dose 1000 times higher to have the same phase error in each pixel.
On the low dose end, we note a peculiar deviation from the error prediction for both ZPC and GS. This occurs because, in both cases, the phase error is bounded. For ZPC, this occurs, e.g., when we get zero counts in a pixel. In that case, the phase is fixed 
at $\phi_i=-10/8$ rad (for 2D), as is obvious from Eq. \ref{eq:z_phase}. Suppose we now calculate the standard deviation of a distribution which is a truncated normal distribution. In that case, we get an RMS value lower than expected from noise considerations only. This leads to a plateau at a very low dose. Note that this plateau does not mean we gained anything regarding information retrieval but is merely an effect of truncation. This effect is clearly visible in a ZPC image of the famous cameraman \cite{Https://hdl.handle.net/1721.3/195767} (see the top rows of Fig. \ref{fig:cameraman}). When using the lowest dose, most pixels are stuck at the lowest value ($-10/8$ for the ZPC case), resulting in a low phase error for an object with limited phase variation. However, these pixels do not contain much information about the object either.

The situation is more complex for the RPI measurements. For a GS implementation, we can think of RPI as a series of $N_{RPI}$ individual measurements. Each of these measurements will have only a fraction of the total dose and, therefore, show a higher error individually. The truncation described above now happens for each realization individually, leading to a phase error plateau lower by $\sqrt{N_{RPI}}$ for the low-dose cases. 
Naturally, this does not happen for the ML implementation because there is only one model with N phases that are bound between $-\pi$ and $+\pi$, and the $N\times N_{RPI}$ detected intensities are correctly dealt with through the likelihood function, which does not suffer from truncation issues.
In order to get the most accurate recorded intensities, it is best to avoid working with dose levels that lead to this plateau. If we need to use GS for computational efficiency, we can do so by choosing $N_{RPI}=2$ for low-dose cases. Another option is to use ML and a higher value of $N_{RPI}$ to make full use of the available information. However, this will require longer computation time and doesn't lead to a significant error reduction.
Note that the plateau is also misleading here, as it could lead to the assumption that the lower plateau level for higher $N_{RPI}$ is a desirable suppression of the noise. However, it is merely a truncation or phase-wrapping artifact that obfuscates the actual signal.

When we examine the visual output of the retrieved phase for the cameraman object in Fig. \ref{fig:cameraman}, we can clearly see some significant differences between ZPC and RPI. At the highest dose, both methods accurately retrieve all object details. In the case of ZPC, the intensity is dispersed outside the illuminated region of interest, which makes it less effective than the diffraction-based method. We discovered that only $62.5\%$ of the intensity is within the region of interest, with the remaining $37.5\%$ not contributing useful information from the object. At low doses (marked with a $\star$ in Fig. \ref{fig:cameraman}), the ZPC displays a flat phase of $-10/8$ \textit{rad} wherever no electron was detected. While this flat value may seem like low noise due to the absence of contrast variation, in reality, it indicates a lack of information in those areas. This is why we observe a plateau in Fig. \ref{fig:error64x64} for ZPC. In the case of GS RPI, this effect also arises from phase wrapping of the noise, but it is less prominent, which is also visible in the figure indicated with a $\star$ in Fig. \ref{fig:cameraman} for RPI.

\begin{figure}
    \centering
    \includegraphics[width=.48\linewidth]{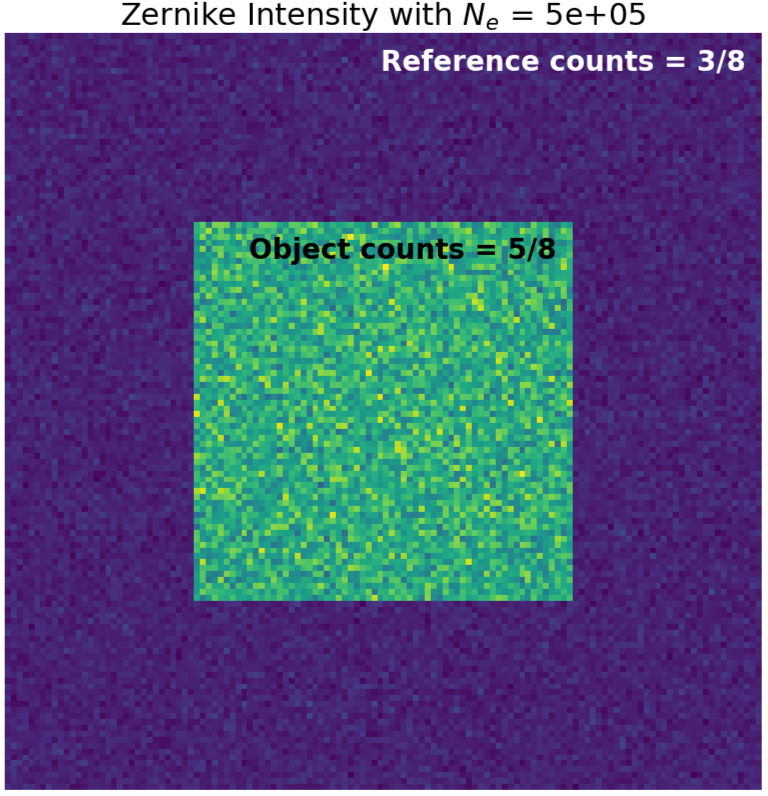}
    \caption{Example of a correctly sampled ZPC image of a random phase object of $64 \times 64$ zero-padded to $128 \times 128$ with $N_e=5\times 10^5$. The central $64 \times 64$ area contains the actual phase-related image and 5/8 of the total dose with which the object was illuminated. The other 3/8 lands outside the central area and does not contribute to the desired information.}
    \label{fig:fiveeights}
\end{figure}

Furthermore, we also show the detected intensity for the ZPC case to demonstrate the normalization effect that was derived in this paper. Fig. \ref{fig:fiveeights} shows the detected intensity of the random phase object of $64 \times 64$. As mentioned before, with correct oversampling of $C=1/2$, we notice the intensity outside of the illuminated area to be 37.5\% of the total illumination intensity. This effect occurs as a result of the phase discontinuity that is essential for ZPC to work but leads to a decrease in dose efficiency as only a fraction of all electrons take part in the actual formation of the central phase encoded part of the image and, even there, the contrast is lower. Omitting this oversampling by 2 in all directions will wrongly create an aliasing of the reference part of the wave, which will result in a seemingly $50/50$ distribution between the object and the reference part of the wave. This distribution will result in much better counting statistics, but it is an aliasing artifact that can not be reproduced in an actual experiment.

\section{Discussion}

Following the previous derivation and results, we can now extract common features from these numerical exercises and discuss the effect of each parameter on the reconstruction.

\subsection{The role of dose}

The role of dose clearly follows the expected $1/\sqrt{N_e/N}$ trend, albeit with different pre-factors for either ZPC or diffraction-based measurements. This trend shows that more electrons are obviously better and that the number of detector pixels $M$ should be carefully balanced against the required sampling and desired field of view. We observe a deviating trend at a very low dose due to phase wrapping, which truncates the phase error and leads to a plateau in the error prediction. This plateau indicates a situation where noise is maximized, and it is unlikely that one would recover any meaningful signal in this range. We observed that this truncation effect is more severe for the GS RPI method as it acts on the $N_{RPI}$ individual experiments, while this does not appear for the ML RPI method. This finding is significant, as it guides the experimenter to use a lower $N_{RPI}$ for very low-dose imaging to optimally distribute the available dose among each phase configuration in the illumination. Ideally, ML RPI would always be used, but our current implementation is too slow to be practical on realistically sized objects. Further algorithm development in this direction would be useful.
As a side note, we speculate that these effects also occur for ptychography with overlapping probes, as also here, one records multiple instances of the same part of the object with different phase encoding \cite{li2014ptychographic,o2020phase,o2021contrast,Leidl2024InfluenceFlow}.

\subsection{Is RPI the best we can do?}

In our study, we employed the RPI method to solve the issue of under-determination in the inverse problem, which helps eliminate translation and point symmetry uncertainties \cite{Fannjiang2012PhaseIllumination}. Although this method provides a five-fold improvement over ZPC, it raises the question of whether it would be more efficient to devise a new illumination scheme based on insights gained from previous experiments rather than recording 16 random phase illumination variants. We explored variations of this approach, such as using the complex conjugate of the current phase estimate, employing orthogonal basis sets like Hadamard and Fourier, and using only $\pm \pi$ phase illumination akin to a charge flipping algorithm \cite{Fienup1982PhaseComparison}. However, these attempts yielded nearly identical results to our simpler random phase illumination scheme.

It is intriguing to consider how we can efficiently encode a phase message using $N_e$ electrons. Assuming that each electron represents one bit of information, then $N_e$ electrons could convey a message with $2^{N_e}$ variations. If we distribute this across the $N=(n \times n)$ independent pixels, we get:
\begin{equation}
\sigma_{\phi,binary}=\frac{2\pi}{2^\frac{N_e}{n^2}}
\end{equation}
This encoding would be significantly more efficient in terms of electron dose, but assumes that:
\begin{itemize}
    \item{We know exactly when an electron is interacting with the sample in order to encode a zero bit when no detecting occurs.}
    \item{We have a way to control the phase of the incoming electron wave precisely.}
    \item{We have a way to set up an experiment to find out if the modulo of the phase shift of the sample with respect to the local incoming phase of the wave is higher or lower than some value.}
\end{itemize}

This scheme could be approached when multiple passes of the same electron through the sample can be configured as proposed in the quantum microscope \cite{Kruit2016DesignsMicroscope,juffmann2017multi,koppell2019design,Turner2021Interaction-FreeElectrons} and even further improvement is expected when using \textit{squeezed} or entangled electron states \cite{koppell2022transmission}. The practical implementation of this seems currently highly challenging. Still, it is good to keep in mind that the results in this paper may differ from what is possible.

\subsection{Technological difficulties to realize this}

In this paper, we have looked at the fundamental counting noise limits and have assumed that we can create an ideal phase plate acting on the phase of the illuminating wave and an ideal but correctly normalized version of the ZPC that acts on the exit wave in the back focal plane. Several technological obstacles have to be overcome in order to realize these ultimate predictions for the ZPC case:
\begin{itemize}
\item{The ideal Airy disc-like Zernike phase profile is difficult to realize.}
\item{The ideal Zernike phase shift may differ from $\pi/2$ (see Bellegia \cite{Beleggia2008AMode}).}
\item{The ZPP profile should adapt in width to the illumination size to obtain the best performance while avoiding loss of low frequencies.}
\end{itemize}
Also, for the RPI, several obstacles will hinder a straightforward implementation:
\begin{itemize}
\item{A programmable phase plate has considerable limitations regarding the fill factor \cite{Yu2023QuantumElectrons,VegaIbanez2023CanTEM,Verbeeck2018DemonstrationElectrons}, meaning that we also imprint amplitude modulation into the probe. However, this may or may not be beneficial in some cases \cite{Allars2021EfficientDiffuser,you2023lorentz}. This could be overcome by scanning the amplitude and phase-modulated probe in ptychography style to illuminate all the areas of the sample with some overlap between the probe positions.}
\end{itemize}
In both scenarios, charging problems, gradual drift, and contamination are potential issues.

\subsection{A note on re-normalisation}

In our paper, we have examined a sample as a collection of random phases without making any assumptions about the relationship between pixels or the probability of certain patterns occurring. In reality, the set of actual images is smaller than the total number of potential random phase formations. In other words, there is some underlying regularity in the sample. If we correctly describe this regularity, we can significantly gain in terms of signal-to-noise as this constitutes prior knowledge about the sample \cite{Schloz2020OvercomingOptimization, Maiden2017FurtherEngine,Maiden2024WASP:Retrieval,Leidl2024InfluenceFlow,diederichs2024exact}. However, we need to be careful and compare similar situations. For example, if we use regularization in an RPI setup, we should use the same assumptions in the case of ZPC. Failing to do so will inevitably lead to biased opinions, as was the case in compressed sensing \cite{donoho2006compressed,leary2013compressed,li2018compressed}, which also relies on regularization arguments. This makes it impossible to compare to a situation where normal sampling is applied unless the regularization prior is included \cite{VandenBroek2019VariousIllumination}. For this reason, we will not discuss regularization here. However, significant gains can be achieved if such prior knowledge is available and valid \cite{diederichs2024exact}.

\section{Conclusion}

We have demonstrated that there is a clear benefit in terms of counting statistics in recording multiple diffraction patterns over the more conventional ZPC imaging.  We stress here that we have shown just one possible way of improving the dose efficiency of ZPC imaging without making claims on related ptychographic or other diffraction-based setups. Nevertheless, breaking this limit for one case demonstrates that ZPC is not the ultimate limit, and this should invite further research. Note also that we have assumed the weak phase object approximation throughout, which is known to be a very limiting assumption in practice. We found the benefit in terms of dose efficiency to be a factor of 5, which would allow a very significant shifting of the beam damage boundaries that are hindering progress in, e.g., life sciences, but also in many materials science areas such as battery materials, polymer science, zeolites, perovskites, metal-organic frameworks and many more.

This improvement factor is mainly attributed to a subtle normalization issue that occurs in ZPC imaging. This issue makes ZPC less dose efficient, as it seems to be the case if we make a fair comparison where the sample is illuminated with exactly the same dose over exactly the same area of illumination.
This paper avoids discussing actual implementation details, and it might well be possible to improve this benefit further if, e.g., the illumination can be updated to take into account information from the partial experiment that was already performed. 
Whether the gain from the diffraction-based recordings proposed here can also be obtained in practice remains to be seen, as many practical details will influence the actual dose efficiency. Nevertheless, there has been much promising progress in ptychography over the last few years, and it relies on diffraction-based detection, albeit with a different illumination scheme than the RPI suggested here.

One potential advantage of detecting in the diffraction plane while adjusting the phase of the probe is the potential for a simpler, smaller, and more cost-effective electron microscope specifically designed for life science imaging. In this scenario, the camera could have significantly fewer pixels, and the projector system could be completely removed (along with the image corrector), as long as some form of scanning system or phase plate can quickly alter the illumination. This discovery shows promise for affordable tabletop instruments that could expand the information we can gather from beam-sensitive nanoscale objects.

\section{Acknowledgements}

J.V. and F.V. acknowledge funding from the Flemish Science Fund, FWO G042820N 'Exploring adaptive optics in transmission electron microscopy.' J.V. acknowledges funding through 
the IMPRESS project, which received funding from the HORIZON EUROPE framework program for research and innovation under grant agreement n. 101094299.

\section{Competing interest}

J.V. declares to have a financial interest in AdaptEM BV \cite{https://adaptem.eu}, a spinoff company from the University of Antwerp that produces adaptive phase plates for the EM market.

\section{Author contributions}

F.V. performed coding of the GS algorithm, numerical exercises, creation of figures, and general writing. J.V. performed derivations in appendices, general conceptualization, coding of ML 1D and 2D simulations, creation of figures, and general writing.

%\begin{appendices}
\appendix

\section{Zernike contrast with proper normalization}
\label{ap:appzernike}
Assume a circular area of illumination on a sample with radius $R$. 
\begin{eqnarray}
M(r)=\Pi\left(\frac{r}{2R}\right)
\end{eqnarray}
With $\Pi(x)$ the Heavyside Pi step function (or rect function) being one for $|x|<\frac{1}{2}$. 
The sample transmission function can be approximated as a phase object:
\begin{eqnarray}
O(\vec{r})=e^{\iu \phi(\vec{r})}
\end{eqnarray}
In the back focal plane of the sample, this will result in
\begin{eqnarray}
\Psi(\vec{k})=\tilde{M}(\vec{k}) \otimes \tilde{O}(\vec{k})=\frac{2J_1(kR)}{kR} \otimes \tilde{O}(\vec{k})
\end{eqnarray}
With $\tilde{}$ designating the Fourier transformed function and $J_1$ a Bessel function of the first kind. 
Typically, experimental realizations of a ZPP in the TEM, such as the Volta Phase Plate, have an Airy-like smooth phase profile \cite{danev2014volta,danev2017using,malac2021phase}. However, to create an idealized ZPP, we assume a maximum passband frequency $k_{max}$, representing the ultimate spatial resolution, and a minimum frequency $k_Z$, below which we introduce the required phase shift of $\pi/2$, modeled as a Heavyside Pi step function as well. A genuinely ideal ZPP will have $k_Z \xrightarrow{} 0 $ and $k_{max} \xrightarrow{} \infty$. However, we will see later that these parameters play an important role in normalization and are critical for discussing quantum information.
\begin{eqnarray}
Z(\vec{k})=\Pi\left(\frac{k}{2k_{max}}\right)\exp\left[\iu \frac{\pi}{2}\Pi\left(\frac{k}{2k_{z}}\right)\right]
\end{eqnarray}
For $k_{max}>k_z$ and we can rewrite this as:
\begin{eqnarray}
Z(\vec{k})=\Pi\left(\frac{k}{2k_{max}}\right) + \Pi\left(\frac{k}{2k_{z}}\right)  (\iu -1) 
\end{eqnarray}
Applying this phase plate in the back focal plane leads to:
\begin{eqnarray}
\Psi_Z(\vec{k})=\left[\frac{2J_1(kR)}{kR} \otimes \tilde{O}(\vec{k}) \right] Z(\vec{k})
\end{eqnarray}
Transforming this back to the image plane, we get:
\begin{eqnarray}
\tilde{\Psi}_Z(\vec{r})=[M(\vec{r}) O(\vec{r})]  \otimes \tilde{Z}(\vec{r})
\end{eqnarray}
We can rewrite the Fourier transform of the ZPP as:
\begin{eqnarray}
\tilde{Z}(\vec{r})=\frac{2J_1(rk_{max})}{rk_{max}} + \frac{2J_1(rk_{z})}{rk_{z}}(\iu-1)
\end{eqnarray}
The first term will lead to a replica of the object with a spatial resolution limit of $1/k_{max}$.
The second term, however, will create a blurred version of the illuminated patch $M$ by a Bessel function of width $R_Z=1/k_Z$. This will serve as a reference wave, revealing the phase contrast when interfering with the first term.
The blurring will take intensity from within the illuminated area outside the illuminated area, which carries no useful information. This is critical for consideration of the role of counting noise, as not all electrons will contribute to the desired ZPC signal. 
The blurring reduces the local intensity by a factor C, which can be estimated in the center of the image as:
\begin{eqnarray}
C&=&\int_0^{Rk_z} \frac{2J_1(k)}{k}  dk\\
&=&[\pi Rk_z H_0(Rk_z) - 2] J_1(Rk_z) \\
&&+ Rk_z [2 - \pi H_1(Rk_z)] J_0(Rk_z)\\
C&\approx&Rk_z 
\end{eqnarray}
With $H_0$ and $H_1$ Struve functions of zeroth and first order. The simplification in the last step is a very good approximation when $Rk_z<1$. If we choose $R k_z=R/R_z \leq 1/2$ (required to recover all but the DC frequency, anything higher will lead to loss of low frequency components) we can rewrite the exit wave for $r \leq R$ as
\begin{eqnarray}
\Psi(\vec{r},r \leq R, R/R_z \leq 1/2 )=O(\vec{r}) +(\iu-1)C^2 \braket{M(\vec{r})O(\vec{r})}
\end{eqnarray}
With $C\approx R/R_z$ and $\braket{M(\vec{r})O(\vec{r})}$ the averaged exit wave inside the illuminated patch $M$ caused by the convolution of the Heavyside and Airy pattern with 'radius' $R$ and $R_z$, respectively. This convolution will re-scale this reference wave by a factor $C^2$ because part of this wave now ends up outside the illumination patch in the image plane.
Applying the weak phase approximation and assuming the phase of the object does not contain spatial frequency variations above $k_{max}$:
\begin{eqnarray}
O(\vec{r}) &\approx& 1+ \iu \phi(\vec{r})\\
\Psi(\vec{r},r\leq R, R/R_z\leq 1/2) &\approx& 1+\iu \phi(\vec{r}) +(\iu-1)C^2\\
I_z(\vec{r},r\leq R, R/R_z\leq 1/2) &\approx& \left|\Psi(\vec{r},r\leq R,C \leq 1/2)\right|^2\\ \nonumber
&\approx& (1-C^2)^2+(\phi(\vec{r}) + C^2)^2\\ \nonumber
&\approx& 1-2C^2+2C^4+2C^2\phi(\vec{r})\\ \nonumber
&&+\phi(\vec{r})^2\\ \nonumber
&\approx& 1-2C^2+2C^4+2C^2\phi(\vec{r})\\ 
&&+\ldots \nonumber
 \label{eq:zernikelinear}
\end{eqnarray}
Where we kept only linear terms in the phase.
$I_z(\vec{r},r \leq R, C \leq 1/2 )$ is the observed intensity in the image plane within the illuminated area. The average intensity within the illuminated patch is $\overline{I_z(\vec{r},r \leq R,C \leq 1/2) }\approx 1-2C^2+2C^4$.
We can recover the phase from the intensity as:
\begin{eqnarray}
\phi(\vec{r},r \leq R) \approx \frac{I_z(\vec{r},r \leq R)-1+2C^2-2C^4}{2C^2}
\label{eq:zernikephasecorrected}
\end{eqnarray}
In a practical implementation, the intensity will be recorded on a pixelated camera, and we assume $I_z$ to be scaled such that if the ZPP was removed, there would be an average of 1 electron per pixel.
This allows us to estimate the standard deviation of the phase error as:
\begin{eqnarray}
\sigma_\phi=\frac{1}{2\sqrt{N_e/n^2}}\frac{\sqrt{1-2C^2+2C^4}}{\vphantom{\hat{f}}C^2}
\end{eqnarray}
Where the last term is a correction term that takes care of the role of $C$, which depends on the parameters of both illumination size and scale of the phase shifting part of the phase plate.

\section{Numerical implementation of Zernike contrast}

\label{ap:appzernikenumerical}
The consequence of the above considerations for a numerical implementation of a ZPP is that zero padding around the object is essential.
Omitting this step will lead to an overly optimistic result on the quantum information content that a ZPC implementation can obtain.
Furthermore, this zero-padding step represents a situation where the object is periodically repeated until infinity and illuminated by an infinite plane wave. In general, we are not interested in perfectly periodic structures (otherwise, diffraction would be a far more logical choice), and even if we are, it is, in practice, impossible to make the illumination box exactly fit with periodic boundaries.

As the intensity recording involves a modulus squared, which relates to the autocorrelation in the image plane, proper zero padding is only guaranteed when there is at least padding with $n/2$ on all sides for an $n\times n$ illumination area.
Padding more will worsen the counting statistics further; padding less will lead to an unphysical situation as it cannot be replicated in a real experiment.

\begin{figure}
    \centering
    \includegraphics[width=.88\linewidth]{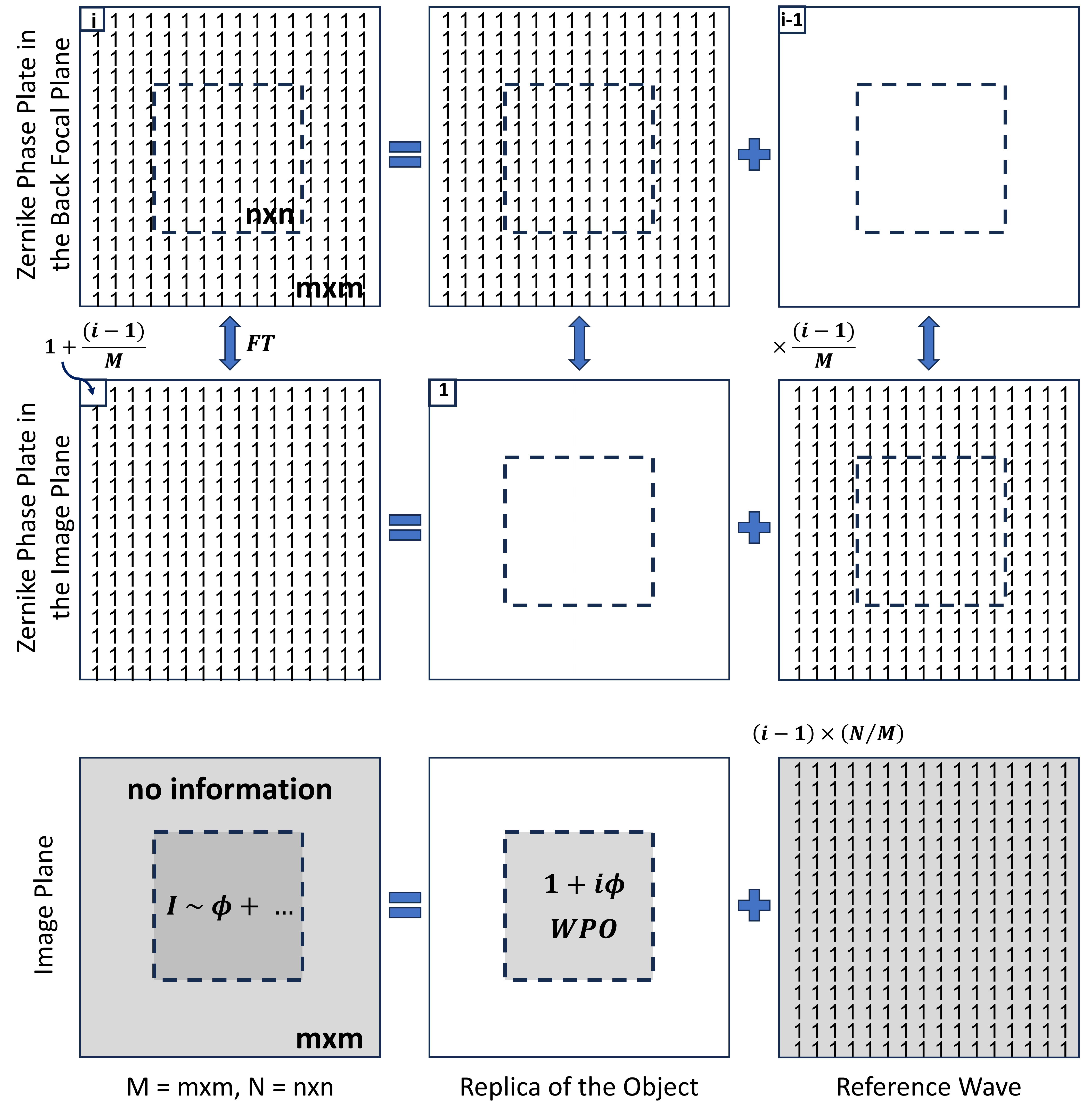}
    \caption{Sketch of a numerical 2D implementation of ZPC imaging that avoids aliasing effects that would lead to errors in estimating the quantum efficiency of the ZPC imaging process.}
    \label{fig:numerical}
\end{figure}

We can use the crudest approximation of a ZPP in a numerical experiment by creating an $M=m\times m$ filter matrix in the discrete Fourier plane with all 1s and only the DC component in the top left corner equal to the complex $\iu$.
We can decompose this filter into a sum of a matrix with all 1s and another matrix with all zeros except for the DC component, which we take as $(\iu-1)$.
Fourier transforming this will lead to a sum of 2 matrices, one of which has only a DC component equal to 1 and the other of which is $1$, and the other being a flat matrix with elements $(\iu-1)/M$.
We can now simulate the action of this filter by convolving this real space representation of the Zernike filter kernel with an object.
If we choose the object first to be the same size as the filter kernel ($M=N$) we get:
\begin{eqnarray}
\Psi_Z(\vec{r})&=& O(\vec{r}) +M \frac{(\iu-1)}{M} <O(\vec{r})>\\
\Psi_Z(\vec{r})&= &1+\iu\phi(\vec{r}) + \iu-1\\
\Psi_Z(\vec{r})&=& \iu+\iu\phi(\vec{r})\\
I_Z(\vec{r})&=& 1+2\phi(\vec{r})+\phi(\vec{r})^2\\
I_Z(\vec{r})&\approx& 1+2\phi(\vec{r})
\end{eqnarray}
From which we can derive the phase unambiguously from the intensity:
\begin{eqnarray}
\phi(\vec{r}) \approx \frac{I_Z-1}{2}
\end{eqnarray}
And this looks exactly like we would expect from a ZPP.
The standard deviation on the phase now becomes:
\begin{eqnarray}
std(\phi(\vec{r}))= \frac{1}{2\sqrt{N_e/M^2}}
\end{eqnarray}
This result is why ZPC is believed to outperform diffraction-based methods, e.g., in \cite{Dwyer2023MaximizingMicroscopy}.

If, on the other hand, we choose to limit the illumination of the object only to fill $N=n\times n$ pixels, we get a different result with $C=\sqrt{N/M}$, an oversampling factor:
\begin{eqnarray}
\Psi_Z(\vec{r})&=&O(\vec{r}) +C^2(\iu-1) <O(\vec{r})>\\
\Psi_Z(\vec{r})&=& 1+\iu \phi(\vec{r}) +C^2 (\iu-1)\\
\Psi_Z(\vec{r})&=& 1-C^2  +\iu (C^2+\phi(\vec{r}))\\
I_Z(\vec{r})&= &1-2C^2+C^4+
 C^4 +2C^2\phi(\vec{r}) + \phi(\vec{r})^2\\
I_Z(\vec{r})& \approx &  1-2C^2+2C^4+2C^2  \phi(\vec{r})
\end{eqnarray}
We can still derive the phase as follows:
\begin{eqnarray}
\phi(\vec{r}) \approx \frac{ I_Z-1+2C^2 -2C^4}{2C^2}
\end{eqnarray}
Note that this is subtly different in several ways.
If $M \geq 4N$, the action of the averaging kernel now smears the object out over the whole matrix but does not cross the borders of that matrix and, therefore, does not cause aliasing. In physics terms, this means that the average wave, which will be used as a reference beam to reveal the phase, now only comes from the illuminated $n\times n$ patch and not from neighboring sample areas, which, in the case of $N=M$, are wrongly assumed to be periodic. Even if we can live with the fact that, for the averaging, it does not matter that the sample is perfectly periodic, we are still using electrons from neighboring areas (and they will cause damage there). On top of this, it will spread the averaged reference wave over a wider area in the image plane (here, $m \times m$ but only if $m \geq 2n$). This area is larger than the $n\times n$ illuminated image region of interest. Therefore, a significant fraction of electrons will end up in areas of the image plane where they do not bring information on the illuminated sample area.

We are now in the position to estimate what fraction of the illumination contributes to the ZPC phase signal.
Outside and inside the illuminated area, we get an average intensity of:
\begin{eqnarray}
\tilde{I}_{Z,out}& \approx&  2C^4\\
\tilde{I}_{Z,in}\; &\approx & 1-2C^2+2C^4
\end{eqnarray}
For an oversampling of $C=1/2$, the average intensity is $1-2/4+1/8=5/8$ inside and $3/8$ on the outside. All intensity is now spread over 4 times more area in real space than before, and only the central quarter will contain information on the sample phase. That central patch only contains $5/8$ of the illumination, but even that $5/8$ contributes less than in the aliased (and therefore wrong) ZPC case because the strength of the averaged reference beam is also reduced.
For 1D, we have $N=n$, $M=m$, and C=1/2. This would be $1/2$ inside and $1/2$ outside, leading to half the intensity, which does not contribute to the area of interest. 

The standard deviation on the phase now becomes:
\begin{eqnarray}
std(\phi_{Z,aliased})&=& \frac{1}{2 \sqrt{N_e/N}}\\
\frac{std(\phi_{Z,2D})}{std(\phi_{aliased})}&=& \frac{\sqrt{1-2C^2+2C^4}}{ C^2} \\
\frac{std(\phi_{Z,1D})}{std(\phi_{aliased})}&=& \frac{\sqrt{1-2C+2C^2}}{ C} 
\end{eqnarray}
For the minimal required oversampling of $C=1/2$, we get
\begin{eqnarray}
\frac{std(\phi_{Z,2D})}{std(\phi_{aliased})}&=& \sqrt{10}\\
\frac{std(\phi_{Z,1D})}{std(\phi_{aliased})}&=& \sqrt{2}
\end{eqnarray}
Identical to the real space derivation as it should be.

\section{Maximum Likelihood model under Poisson noise}
\label{ap:appmodelbased}
The parameter-dependent part of the log-likelihood for Poisson is given as:
\begin{equation}
l=  -\sum_l^N (I_{exp,l} \ln(I_{model,l})- I_{model,l})
\end{equation}
With $I_{exp,l}$ the experimental observation and $I_{model,l}$, the model prediction is based on a set of detector pixel $l$ parameters.
The model is obtained as follows:
\begin{eqnarray}
\Psi_i&=&A_i e^{\iu (\phi_i+\alpha_i)}\\
I_{model,l}&=&\left| \mathcal{F}_l A_i e^{\iu (\phi_i+\alpha_i)}\right|^2\\
I_{model,l}&=&\left| \sum_{i} A_i e^{\iu (\phi_i+\alpha_i)}e^{2 \pi \iu \vec{r_i} \cdot \vec{k_l}}\right|^2=| \tilde{\Psi}_l|^2
\end{eqnarray}
We included the optional illumination function $A_i e^{\iu \alpha_i}$, allowing for both an amplitude and phase-modulated illumination pattern.

Differentiating the log-likelihood from the parameters of the model gives:
\begin{eqnarray}
\frac{\partial l}{\partial \phi_i} =  -\sum_l  \left(\frac{I_{exp,l}}{I_{model,l}}- 1\right) \frac{\partial I_{model,l}}{\partial \phi_i}
\end{eqnarray}
Differentiating the model intensities from the model parameters, we get the:
\begin{eqnarray}
\frac{\partial I_{model,l}}{\partial \phi_i}&=&-2 A_i \textbf{Re}{\tilde{\Psi}} \sin(\phi_i+\alpha_i+r_ik_l)
\nonumber\\
&&+2 A_i\textbf{Im}{\tilde{\Psi}}\cos(\phi_i+\alpha_i+r_ik_l)
\end{eqnarray}
Suppose we constrain the last phase value $\phi_N$ to maintain an average phase of zero (the absolute phase has no meaning, and fitters do not like duplicate parameters). In that case, we need to take into account this dependent variable. So, we proceed now with $(N-1)$ independent parameters:
\begin{eqnarray}
\frac{\partial I_{model,l}}{\partial \phi_i}&=&-2 \textbf{Re}{\tilde{\Psi}_l} [A_i\sin(\phi_i+r_ik_l+\alpha_i)\nonumber\\
&&-A_N\sin(\phi_n+r_n k_l+\alpha_n)]\nonumber\\
&&+2\textbf{Im}{\tilde{\Psi}_l}[A_i\cos(\phi_i+r_ik_l+\alpha_i)\nonumber\\
&&-A_N\cos(\phi_n+r_n k_l+\alpha_n)]
\end{eqnarray}
Using this analytical derivative as a Jacobian input for the nonlinear fitter significantly speeds up the process and improves convergence.
We can now derive the Fisher information matrix:
\begin{eqnarray}
F_{i,j} &=& -E\left[\frac{\partial^2 l}{\partial \phi_i\partial \phi_j}\right]\\
&=&  \sum_l  \frac{\partial}{\partial \phi_j}\left(\frac{I_{exp,l}}{I_{model,l}}- 1\right) \frac{\partial I_{model,l}}{\partial \phi_i}\nonumber\\
&&+\left(\frac{I_{exp,l}}{I_{model,l}}- 1\right) \frac{\partial I_{model,l}}{\partial^2 \phi_j\partial \phi_i}\\
&=&- \sum_j \frac{1}{I_{model,l}} \frac{\partial I_{model,l}}{\partial \phi_i}\frac{\partial I_{model,l}}{\partial \phi_j}
\end{eqnarray}
From this, we can obtain the $(N-1)\times(N-1)$ covariance matrix:
\begin{eqnarray}
 cov=F^{-1}
\end{eqnarray}
From which the diagonal elements represent the CRLB of the variance on the individual phase estimates $\tilde{\phi_i}$. This allows us to write a lower limit for the standard deviation on the estimated parameters as: 
\begin{eqnarray}
\sigma_{ML} \geq \sqrt{\frac{\textbf{Tr}(cov)}{N-1}}
\end{eqnarray}

\bibliographystyle{unsrt}
\bibliography{RPI}

\end{document}